\documentclass[iop]{emulateapj}

\usepackage{graphics,graphicx}
\usepackage{psfig}
\usepackage{subfigure}
\usepackage{appendix}
\usepackage{booktabs}

\providecommand{\mu}{$\Delta m_{15}(U)$}

\begin{document}
\title{The 1999\lowercase{aa}-\uppercase{like}
Type I\lowercase{a} Supernova \lowercase{i}PTF14\lowercase{bdn} in the Ultraviolet and Optical}
\author{Michael~T.~Smitka, Peter~J.~Brown, Nicholas~B.~Suntzeff}
\affil{George P. and Cynthia Woods Mitchell Institute for Fundamental Physics \& Astronomy, \\
Texas A\&M University, Department of Physics and Astronomy, \\
4242 TAMU, College Station, TX 77843, USA }
\author{Jujia Zhang, Qian Zhai}
\affil{Yunnan Observatories (YNAO), Chinese Academy of Sciences, Kunming 650011, China. \\
University of Chinese Academy of Sciences, Chinese Academy of Sciences, Beijing 100049, China. \\
Key Laboratory for the Structure and Evolution of Celestial Objects, Chinese Academy of Sciences, Kunming 650011, China.}
\author{Xiaofeng Wang, Jun Mo}
\affil{Physics Department and Tsinghua Center for Astrophysics (THCA), Tsinghua University, Beijing 100084, China.}
\author{Tianmeng Zhang}
\affil{Key Laboratory of Optical Astronomy, National Astronomical Observatories, Chinese Academy of Sciences, Beijing 100012, China}            

\email{mikesmitka@gmail.com}  

\begin{abstract}

We present ultraviolet (UV) and optical photometry and spectra of the 1999aa-like supernova (SN) iPTF14bdn.  The UV data were observed using the {\sl Swift} Ultraviolet/Optical Telescope (UVOT) and constitute the first UV spectral series of a 1999aa-like SN.  From the photometry we measure $\Delta m_{15}({\it B})\,=\,0.84 \pm0.05$~mag and blue UV colors at epochs earlier than $-5~$ days.  The spectra show that the early-time blue colors are the result of less absorption between $2800~-~3200$~\AA~ than is present in normal SNe Ia.  Using model spectra fits of the data at $-10~$ and $+10~$ days, we identify the origin of this spectral feature to be a temperature effect in which doubly ionized iron group elements create an opacity 'window'.  We determine that the detection of high temperatures and large quantities of iron group elements at early epochs imply the mixing of a high Ni mass into the outer layers of the SN ejecta.  We also identify the source of the {\it I}-band secondary maximum in iPTF14bdn to be the decay of Fe III to Fe II, as is seen in normal SNe Ia.
 
\end{abstract}

\keywords{(stars:) supernovae: individual (iPTF14bdn, SN 1991T, SN 1999aa), ultraviolet: stars, line: formation, line: identification }

\section{Introduction  \label{intro}}

\indent The use of Type Ia supernovae (SNe Ia) as distance indicators revolutionized cosmology by revealing the accelerating universe \citep{Riess1998,Perlmutter1999} and are widely viewed as a valuable tool in efforts to distinguish between differing cosmological models and constraining cosmological parameters $H_o$, $\Omega_m$ and $\Omega_\Lambda$ \citep{Barris2004,Wood-Vasey2007,Amanullah2010,Kessler2009,Sullivan2011}.  The underlying priniciple which makes this possible is that normal SNe Ia are excellent standardizable candles due to a correlation between peak luminosity and light-curve width \citep{Pskovskii1977,Phillips1993}.  This standardizability is tied to the rest-frame optical regime. 

While the rest-frame ultraviolet regime doesn't appear to provide useful standard candles due to greater scatter in UV absolute magnitudes and colors \citep{Jha2006, Brown2010}, it has provided new insights into normal SNe Ia physics, practical means of subclassifying normal SNe Ia and better methods of identifying peculiar SNe Ia than are apparent in the optical.  The finding of \citet{Milne2013} that optically normal SNe can be subclassified into near-UV red and near-UV blue categories exemplifies color differences among SNe which only become apparent when looking in the UV.  More recently \citet{Brown2014a} and \citet{Brown2014b} have shown that with decreasing wavelength into the UV the deviant behavior of some peculiar SNe Ia towards brighter absolute magnitudes and bluer colors increases.  Explanations for this UV excess cite evidence of an interaction of the SN ejecta with a companion star and the possibility of masses in excess of the Chandrasekhar limit \citep{Howell2006,Brown2014a}.  In the case of these super-Chandrasekhar SNe, the overluminous behavior is sometimes only detected in the UV absolute magnitudes and not the optical, thus stressing the importance of understanding the rest-frame UV for peculiar SNe Ia.  As large ground-based surveys like DES and LSST detect SNe Ia at higher and higher redshifts knowledge of the rest-frame UV will be crucial in interpreting their observations.  

Peculiar SNe Ia are classified by how their light curve shapes and spectral features differ from normal SNe Ia.  The peculiar 1991T-like SN subclassification is defined in the optical by overluminous, slowly declining light curves, weak or no Ca II and Si II absorption at epochs earlier than $\sim 1$~week, and strong Fe III absorption \citep{Filippenko1992,Phillips1992}.  The peculiar 1999aa-like SNe classification is similar to 1991T-like with the subtle differences of having weak signatures of Ca II H \& K and Si II absorption prior to maximum light \citep{Garavini2004}.  From optical observations, possible explanations of the peculiar behavior of both classes include abnormal surface abundances, progenitor deflagration or delayed detonation, and higher than normal photospheric temperatures \citep{Phillips1992,Garavini2004,Sasdelli2014}.  Few UV observations exist for these SNe.  \citet{Jeffery1992} presented three near-maximum {\it IUE} spectra of 1991T and found a suitable model where the outer envelope is dominated with newly synthesized Ni and Co.  {\it Swift} UV photometry as early as $-7~$ days of two 1991T-like SNe has shown brighter than normal UV absolute magnitudes, roughly normal color evolution in the NUV and hints of bluer than normal colors in the mid-UV \citep{Brown2014a}.

We present UV and optical photometry and spectral series of iPTF14bdn, a high-luminosity 1999aa-like SN Ia, observed using the Ultraviolet/Optical Telescope (UVOT) \citep{Roming_etal_2005} onboard the {\it Swift} spacecraft \citep{Gehrels_etal_2004} and the Lijiang $2.4$~meter telescope.  The UV data comprise the first published UV photometric and spectral series of a 1999aa-like SN Ia.  
In Section \ref{obs} we present the UV and optical photometry and spectroscopy of iPTF14bdn.  In Section \ref{results} we present light curves, UV colors, compare the UV spectra to those of normal SNe Ia and to the bluest rest-frame spectral observations of 1991T-like SNe and present SYNAPPS ion analysis of the spectra.  In Section \ref{discuss} we discuss the origins of the early-time UV flux and the {\it I}-band secondary maximum.

\section{Observations of \lowercase{i}\uppercase{PTF14}\lowercase{bdn}}  \label{obs}
iPTF14bdn \citep{2014ATel.6175} was discovered in UGC 8503 ($z=0.01558$) \citep{Adelman2007} on 2014 May 27.40 by the intermediate Palomar Transient Factory \citep{Law2009} at a phase of 18 days prior to {\it B}-band maximum light and identified as a 1991T-like or 1999aa-like SN Ia by a weak Si II $6155$ \AA~ feature.  No comment was made on the presence of a Ca II H \& K feature to distinguish between classification as 1991T-like or 1999aa-like in the discovery announcement.  No sign of strong extinction was reported in the discovery spectrum nor observed later by us.  \citet{Schlafly2011} cite a Galactic extinction of $A_{V\_Landolt}~=~0.032$ mag in the direction of UGC8503.

\subsection{{\it Swift} Ultraviolet Observations} \label{swiftobs} 
We began a {\it Swift} UVOT photometric monitoring program on 2014 May 27 and determined that the target had become bright enough for spectroscopic observations to begin on 2014 June 05.  We obtained UVOT photometry in the six photometric bandpasses from phases $-18$ days through $+45$ days relative to B-band maximum light and was analyzed using the reduction method of the {\it Swift} Optical/Ultraviolet Supernova Archive (SOUSA) \citep{Brown2014SOUSA}.  We removed host galaxy light from the photometry in all six bandpasses using template images obtained one year after the initial observations.  The reduced photometry is presented in Table \ref{tablephot}.

Six epochs of spectra were observed between phases $-9$ through $+18$~days using both the UV and V grisms operating in clocked mode.  The UVOT spectra snapshots were extracted, wavelength calibrated and flux calibrated using \uppercase{UVOTPY} \citep{Kuin2014,Kuin2015} and were then coadded for each epoch using variance weighting.  Our extraction process for the UV grism images utilized galaxy template images observed on 2015 May 19 and June 4 to remove light from the host galaxy and other contaminant sources near the SN \citep{Smitka2015}.  This resulted in the removal of a feature near $2500$~\AA~ in the $-6$~ and $-3$~day spectra which we attribute to a faint zero order diffraction source in the grism images.  Details of the spectroscopic observations are presented in Table \ref{table1}.   \footnote {The {\it Swift} UVOT photometric transmission functions, grism profiles, and their associated calibration products are available at http://heasarc.gsfc.nasa.gov/docs/heasarc/caldb/data/swift /uvota/index.html} 

\subsection {Optical Observations}
We observed optical photometry and spectrophotometry with the Lijiang $2.4m$ telescope and the Yunnan Faint Object Spectrograph and Camera (YFOSC) \citep{Wang2008}.  Photometric observations were made in the {\it UBVRI} bandpasses between the dates of 2014 May 29 and 2014 July 24.  We performed background subtraction of the host galaxy light for all filters using template observations gathered on 2015 March 17.  To estimate the residual brightness of iPTF14bdn in the template images we used the {\it UBVRI} photometry of SN 1991T \citep{Lira1998}, which extends $\sim 400$~days past maximum in all bands.  By assuming that iPTF14bdn behaved similarly to SN 1991T in all bands, we found that we should expect the {\it B} band to contain the brightest residual source in which the SN is $\sim 22$~mag ($\sim 0.001$ of the maximum-light flux).  The lack of extinction is suggestive that the SN exploded in a low-dust environment and that we should not expect light echoes to be present in the template images.  
Following template subtraction, we made photometric measurements using aperture photometry.  Spectroscopic observations were made on the dates shown in Table \ref{table1} spanning phases of $-15$ through $+40$~days relative to B-band maximum light.  \footnote{The photometric response curves of the YFOSC instrument are presented in the Appendix.}

Additional photometric measurements in the {\it UBVRI} bandpasses were made using the $80$~cm Tsinghua-NAOC (National Astronomical Observatories of China) Telescope (TNT) between the dates of 2014 June 18 and 2014 June 28.  These data were reduced and processed using the same methods as the YFOSC data.

\begin{center}
\begin{deluxetable}{cccc}
\tablecaption{Photometry of iPTF14bdn \label{tablephot}}
\tablecolumns{6}
\tablehead{\colhead{Filter} & \colhead{MJD} & \colhead{Mag} & \colhead{Mag err} }
\startdata
UVW2 & 56806.51 & 18.96 &  0.19 \\  
UVW2 & 56808.62 & 19.08 &  0.20   \\  
UVW2 & 56812.45 & 18.19 &  0.13   \\  
UVW2 & 56813.51 & 18.11 &  0.12   \\  
UVW2 & 56815.02 & 17.88 &  0.11  \\  
UVW2 & 56816.61 & 18.03 &  0.12   \\  
UVW2 & 56816.69 & 17.82 &  0.11   \\  
UVW2 & 56817.26 & 17.99 &  0.15   \\  
UVW2 & 56819.49 & 17.88 &  0.14   \\  
UVW2 & 56820.55 & 17.78 &  0.11     
\enddata
\tablecomments{This table is available in its entirety in machine-readable form in the online journal.}
\end{deluxetable}
\end{center}

\begin{center}
\begin{deluxetable*}{lcccccc}
\tablecaption{Spectroscopy of iPTF14bdn \label{table1}}
\tablecolumns{6}
\tablehead{\colhead{Date (2014)} & \colhead{MJD +56000} & \colhead{SN Phase} & \colhead{Integration Time} & \colhead{Telescope} & \colhead{Instrument} \\ \colhead{ } & \colhead{ } & \colhead{(days)} & \colhead{(sec)} & \colhead{ } & \colhead{ } }
\startdata
May  \negthinspace 30  & 807.7 &-14.8& 1800 & LiJiang 2.4m & LJT+YFOSC  \\
Jun \thinspace 01   & 809.6 &-12.9& 2100 & LiJiang 2.4m & LJT+YFOSC  \\
Jun \thinspace 03   & 811.7 &-10.8& 1800 & LiJiang 2.4m & LJT+YFOSC  \\
Jun \thinspace 05   & 813.5 &  -9.0  & 6497 & {\it Swift} & UVOT V grism \\
Jun \thinspace 08   & 816.6 &   -5.9  & 7270 & {\it Swift} & UVOT UV grism \\
Jun \thinspace 09   & 817.6 &  -4.9  & 1800 & LiJiang 2.4m & LJT+YFOSC  \\
Jun \thinspace 11 & 819.5 &  -3.0 & 4590 & {\it Swift} & UVOT UV grism \\
Jun \thinspace 19 & 827.6 &5.0    & 9140 & {\it Swift} & UVOT UV grism \\
Jun \thinspace 23 & 831.2 & 8.7    & 8617 & {\it Swift} & UVOT UV grism \\
Jun \thinspace 24 & 832.6 &10.1  & 2100 & LiJiang 2.4m & LJT+YFOSC  \\
Jun \thinspace 27 & 835.3 &12.8  & 7840 & {\it Swift} & UVOT V grism \\
Jul  \thinspace \space 02    & 840.3 &17.8  & 5660 & {\it Swift} & UVOT UV grism \\
Jul  \thinspace \space 05    & 843.6 &21.0  & 2100 & LiJiang 2.4m & LJT+YFOSC \\
Jul  \thinspace \space 24  & 862.6 &40.0  & 2100 & LiJiang 2.4m & LJT+YFOSC

\enddata
\end{deluxetable*}
\end{center}

\begin{figure}
\hspace*{-.4in}
\epsscale{1.25}
\plotone{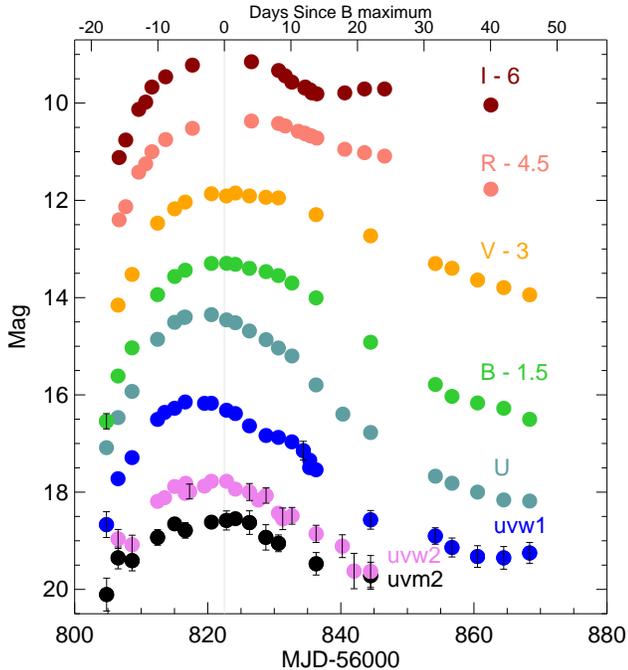}
\caption{Lightcurves of iPTF14bdn.  Data shown here for the {\it UBV} filters are those of UVOT.  Error bars are shown only for data points with photometric errors greater than $0.15$~mag.  The light grey vertical line marks B-band maximum.  Host galaxy light has been removed from all photometry.}
\label{lightcurves}
\end{figure}

\section{Analysis}\label{results}
\subsection{Light Curves} \label{sectionlightcurves}
Light curves of iPTF14bdn are displayed in Figure \ref{lightcurves}.  Visual inspection of the light curves shows very slow declining behavior and an I-band secondary peak about $22$~days after B-band maximum.  To calculate light curve parameters for the UVOT photometry we used a version of the SNooPy light curve fitting algorithm \citep{Burns2011} that had been modified by Chris Burns to accept the UVOT photometric bandpasses and calculate S-corrections \citep{Suntzeff2000} onto the Carnegie Supernova Project photometric system for the {\it u}, {\it B} and {\it V} bandpasses \citep{Hamuy2006,Contreras2010,Stritzinger2011}.  Using this we calculate {\it B}-band maximum light to occur at $MJD\,=\,56822.5 \pm0.3$ and $\Delta m_{15}(B)\,=\,0.84 \pm0.05$~mag.  This slow declining behavior is characteristic of the 1991T-like and 1999aa-like classes of objects and does not provide adequate information for distinguishing between these two categorizations.  Light curve parameters for the ground-based Lijiang 2.4m and TNT photometry were calculated using smooth interpolating spline fits to the data.  Our calculated light curve parameters are presented in Table \ref{tablelcpar}

\begin{center}
\begin{deluxetable}{cccc}
\tablecaption{Light curve parameters of iPTF14bdn \label{tablelcpar}}
\tablecolumns{4}
\tablehead{\colhead{Filter} & \colhead{Peak MJD} & \colhead{Peak Mag} & \colhead{$\Delta m_{15}$} \\ \colhead{ } & \colhead{ } & \colhead{} & \colhead{(mag)} }
\startdata
UVM2 & $56819.6 \pm 0.6$& $18.63 \pm 0.04$& $0.59 \pm 0.21$ \\
UVW2 & $56819.3 \pm 0.3$& $17.84 \pm 0.03$& $0.83 \pm 0.11$ \\   
UVW1 & $56818.4 \pm 0.2$& $16.18 \pm 0.02$& $1.02 \pm 0.05$ \\
U        & $56819.0 \pm 0.2$& $14.28 \pm 0.02$& $1.27 \pm 0.07$\\
B        & $56822.5 \pm 0.3$& $14.75 \pm 0.02 $& $0.84 \pm 0.05$\\
V        & $56823.2 \pm 0.2$& $14.83 \pm 0.01$& $0.60 \pm 0.04$\\
R        & $56825.7 \pm 0.4$& $14.87 \pm 0.02$& $0.61 \pm 0.04$\\
I         & $56822.7 \pm 0.2$& $15.10 \pm 0.02$& $0.67 \pm 0.02$
\enddata 
\end{deluxetable}
\end{center}

\subsection{Photometric Colors}
\begin{figure}
\hspace*{-.4in}
\epsscale{1.35}
\plotone{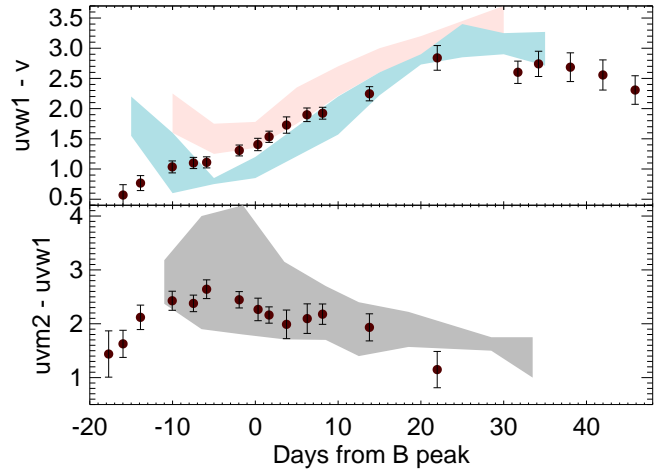}
\caption{The color evolution of iPTF14bdn in the UVOT bandpasses.  The red and blue shaded regions outline the range of colors observed for the evolution of NUV-red and NUV-blue normal SNe Ia respectively, as defined by \citet{Milne2013}.  The gray shaded region outlines the range of colors observed for normal SNe Ia found by \citet{Brown2014a}.  The early-time evolution is shown to contrast normal SNe Ia in both colors and transition towards the normal evolution trend around $5$~days prior to B-band maximum.}
\label{colorplot}
\end{figure}

The color evolution of iPTF14bdn in {\it uvm2 - uvw1} and {\it uvw1 - V} are presented in Figure \ref{colorplot}.  In both colors iPTF14bdn is bluer than normal at the earliest epochs.  Its early time evolution is toward redder colors in contrast to normal SNe Ia at this epoch.  Around $5$~days prior to B-band maximum we observe iPTF14bdn to transition to normal SNe Ia evolution in both colors.   

\subsection{Spectra}
\begin{figure}
\epsscale{1.175}
\plotone{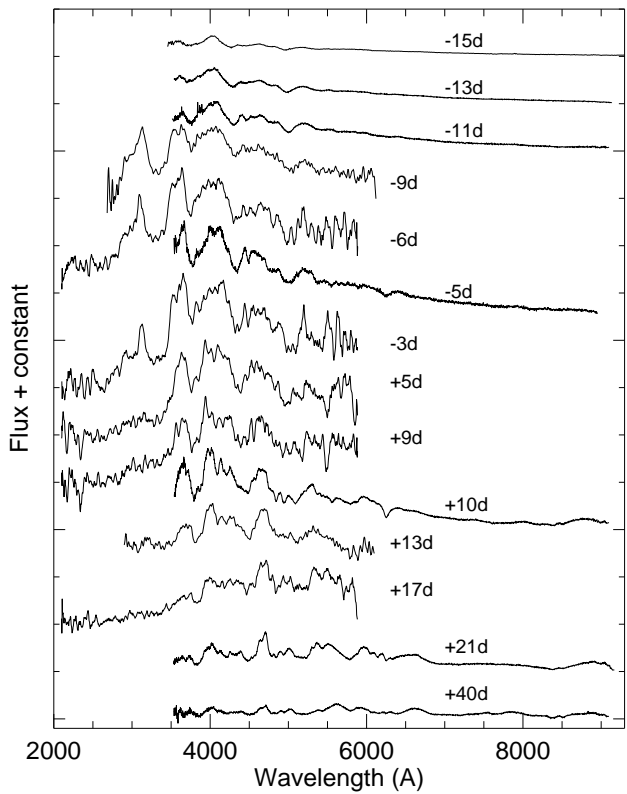}
\caption{Spectra of iPTF14bdn.  Phases shown are calculated relative to B-band maximum light at $MJD\,=\,56822.5 \pm 0.3$.}
\label{spectra1}
\end{figure}

The spectra of iPTF14bdn shown in Figure \ref{spectra1} represent the first published ultraviolet time-series spectra of a 1999aa-like SN Ia.  Upon visual inspection we observe a weak Ca II H \& K feature near $3800$~\AA~ at early times and its subsequent evolution towards normal SN Ia levels starting around maximum light.  This characteristic was first observed by \citet{Filippenko1992} in SN 1991T and later by \citet{Garavini2004} in SN 1999aa.  We see a lack of strong Si II $6355$~\AA~ features prior to maximum light and its subsequent development days after maximum light, also characteristic of 1991T-like and 1999aa-like objects \citep{Filippenko1992,Garavini2004}.

\subsection{Classification}
\begin{figure}
\epsscale{1.15}
\plotone{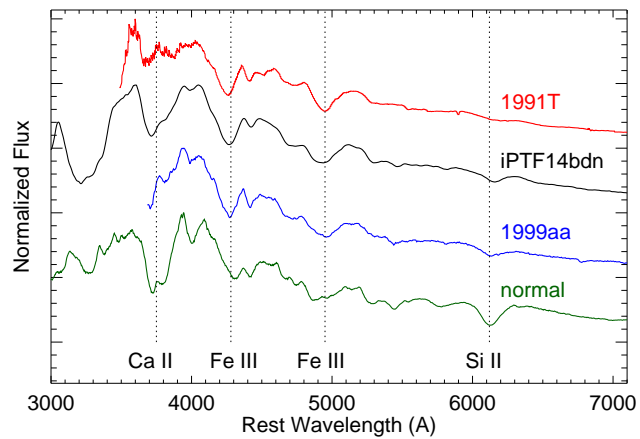}
\caption{We spectroscopically classify iPTF14bdn at $-6$~days by comparison to the spectra shown.  The features of weak Si II, strong Fe III and the detection of Ca II are most consistent with SN 1999aa.  The normal SNe Ia shown is SN 2011fe.  All comparison spectra are from epochs within $1$~day of our spectrum. \citep{Mazzali1995,Garavini2004,Mazzali2014}.}
\label{classification}
\end{figure}

To classify iPTF14bdn we compared the {\it B}-band photometric light curve parameters and $-6~$ day spectrum to those of SNe Ia subclasses.  We find that $\Delta m_{15}(B)\,=\,0.84$~mag of iPTF14bdn is consistent with that of SN 1999aa with $\Delta m_{15}(B)\,=\,0.85$~mag \citep{Jha2002} and SN 1991T with $\Delta m_{15}(B)\,=\,0.95$~mag \citep{Lira1998}.  A spectral comparison of iPTF14bdn at $-6~$ days to SNe 1999aa, 1991T and a normal SNe Ia is shown in Figure \ref{classification}.  We conclude that in the optical iPTF14bdn is a clone of SN 1999aa at this epoch.  

\newpage
\subsection{Comparison to SN 2011fe} \label{comp11fe}
Using abundance tomography modeling, \citet{Sasdelli2014} determined that the optical spectra of SN 1991T at early times displayed several highly ionized features not associated with normal SNe Ia, notably strong Fe III, Co III, Si III, and the lack of strong Si II and Ca II H \& K features.  In the week prior to maximum light the spectra develop more normal Si II features and then evolve to be indistinguishable from normal SNe Ia by one week after maximum light.  \citet{Garavini2004} cites similar findings for SN 1999aa and notes that the resemblance to normal spectra is seen to occur at slightly earlier epochs than SN 1991T.  

\begin{figure}
\epsscale{1.2}
\plotone{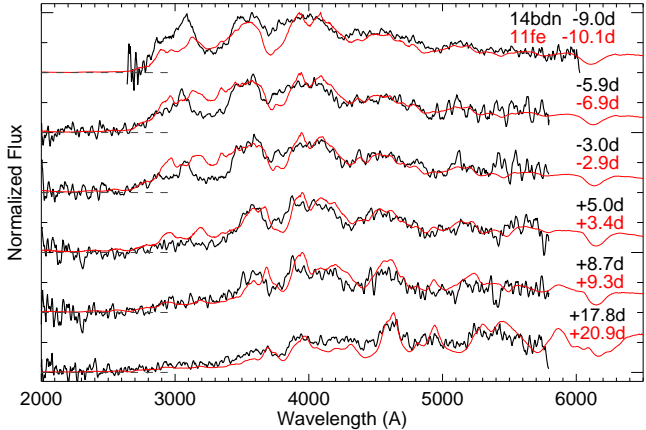}
\caption{Comparison of the rest-frame spectra of SN iPTF14bdn to the normal SN 2011fe.}
\label{spectra3}
\end{figure}

In the ultraviolet we observe an overall similar trend from deviancy toward normalcy upon comparison to SN 2011fe  in Figure \ref{spectra3} \citep{Mazzali2014}.  The Ca II H \& K feature near $3800$~\AA~ is initially very weak at more than a week prior to maximum light and grows progressively more normal with time, becoming indistinguishable between $+5$ and $+9$~days.  This is indicative of a highly energized atmosphere prior to maximum light.

In the $2800~-3200$~\AA~ range we see much less UV absorption than is normal at 9 days prior to maximum light.  This bright feature is shown to evolve to normal flux levels by 6 days prior to maximum light.  Leading up to maximum light we see a significant deficiency of flux in the $3000-3500$~\AA~ range.  The models of SN 1991T of \citet{Sasdelli2014} suggest that this strong absorption feature is attributable to a combination of Fe III, Co II, and Co III indicating a highly energized atmosphere.  
By 5 days following maximum light we observe the evolution blueward of $3500$~\AA~ to be largely indistinguishable from SN 2011fe.

We observe no major differences between SN 2011fe and iPTF14bdn blueward of $2800$~\AA.  In this region both SNe appear to evolve similarly, indicating that line blanketing is playing a major role in suppressing flux.

\subsection{Comparison to SN 2012fr} \label{comp12fr}
\begin{figure}
\epsscale{1.2}
\plotone{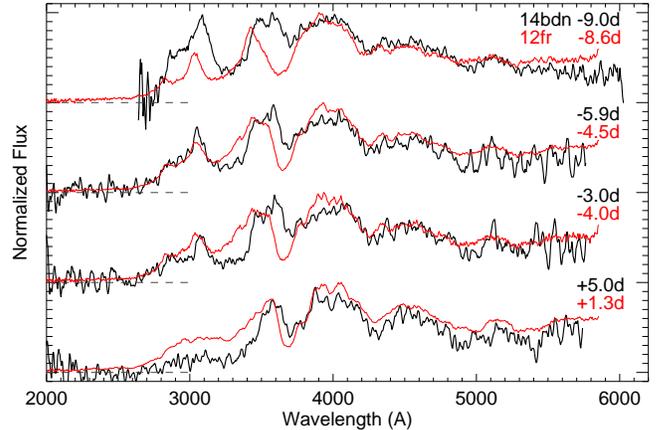}
\caption{Comparison of the rest-frame spectra of iPTF14bdn to UVOT UV-grism spectra of the luminous, slow declining SN 2012fr.}
\label{spectra4}
\end{figure}

Based on a thorough analysis of the optical spectra of SN 2012fr, \citet{Childress2013} concluded that 2012fr was not a 1991T-like or 1999aa-like SN with high velocity features superimposed, but was rather best classified as a very luminous, slow declining normal SN $(\Delta m_{15}(B)=0.80$~mag).  This conclusion was based on the strong presence and high velocity of a Si II $6355$~\AA~ feature and a lack of strong Fe signatures characteristic of 1991T-like SNe.  In a separate analysis, \citet{Zhang2014} presented {\it Swift} UVOT and optical photometry and optical spectra of SN 2012fr.  Their analysis found $\Delta m_{15}(B)=0.85$~mag and concluded that SN 2012fr is likely a member of a subset of 1991T-like SNe Ia whose differences can be the result of viewing angle and asymmetric ejecta.  In Figure \ref{spectra4} we present a comparison of UVOT UV-grism spectra of SN 2012fr and iPTF14bdn.  We extracted and coadded the spectra of SN 2012fr using the same methods as iPTF14bdn.  No galaxy template images were used in the extraction of these spectra because no signs of contamination are present.  A composite maximum light UVOT spectrum of this SN was first presented by \citet{Brown2014c} and the full UVOT spectral series is presented here for the first time.  

Redward of $4000$~\AA~ we note that the spectra appear very similar at all epochs for which we make comparisions.  We see weaker Ca II H \& K absorption in iPTF14bdn than is present in SN 2012fr, as \citet{Childress2013} reported for 1991T-like SNe.

As with SN 2011fe, we see less absorption between $2800-3200$~\AA~ in iPTF14bdn around 9 days prior to maximum light.  We note that the shapes of the features in this region are more similar here than was the case with SN 2011fe.  Continuing to maximum light we also again observe a deficiency of flux in the $3000-3500$~\AA~ range in iPTF14bdn, where we now note that the overall shape of these absorption features are more similar than was the case with SN 2011fe.  Again, we notice no significant differences blueward of $2800$~\AA.

\subsection{SYNAPPS Models of iPTF14bdn} \label{synappssection}
The synthetic spectra synthesis program SYN++ \citep{Thomas2011} is derived from SYNOW \citep{Fisher1997}.  It computes synthetic spectra of SNe Ia in the photospheric phase by modeling the radiative transfer of photons emitted by a sharp photosphere below an expanding ejecta atmosphere.  Spectra lines are assumed to form by resonance scattering using the Sobolev approximation of a moving atmosphere \citep{Sobolev1960,Jeffery1989}.  The SYNAPPS code automates a $\chi ^2$ minimization routine which identifies the combination of SYN++ model atmosphere parameters that provide the best matching synthetic spectrum to an observed spectrum.  The SYN++ and SYNAPPS fitting parameters include the photospheric temperature and velocity, and individual ion temperatures, velocities, ionization states, and optical depths.  The program does not distinguish between stable and radioactive atomic species and does not simulate isotopic line shifts.  We choose to focus only on line identifications made by the SYN++ and SYNAPPS programs and their analogous features in our observed spectra.  It is noteworthy that SYNAPPS is not an ab initio SNe Ia modeling code and so does not offer as much insight into the SN progenitor system and explosion mechanism as other methods \citep{Woosley2011}.  Thus, we encourage additional analysis of the data we present here using other modeling techniques.

\begin{figure*}[ht]
\centering
\subfigure{%
\includegraphics[height=2.8in,width=3.4in]{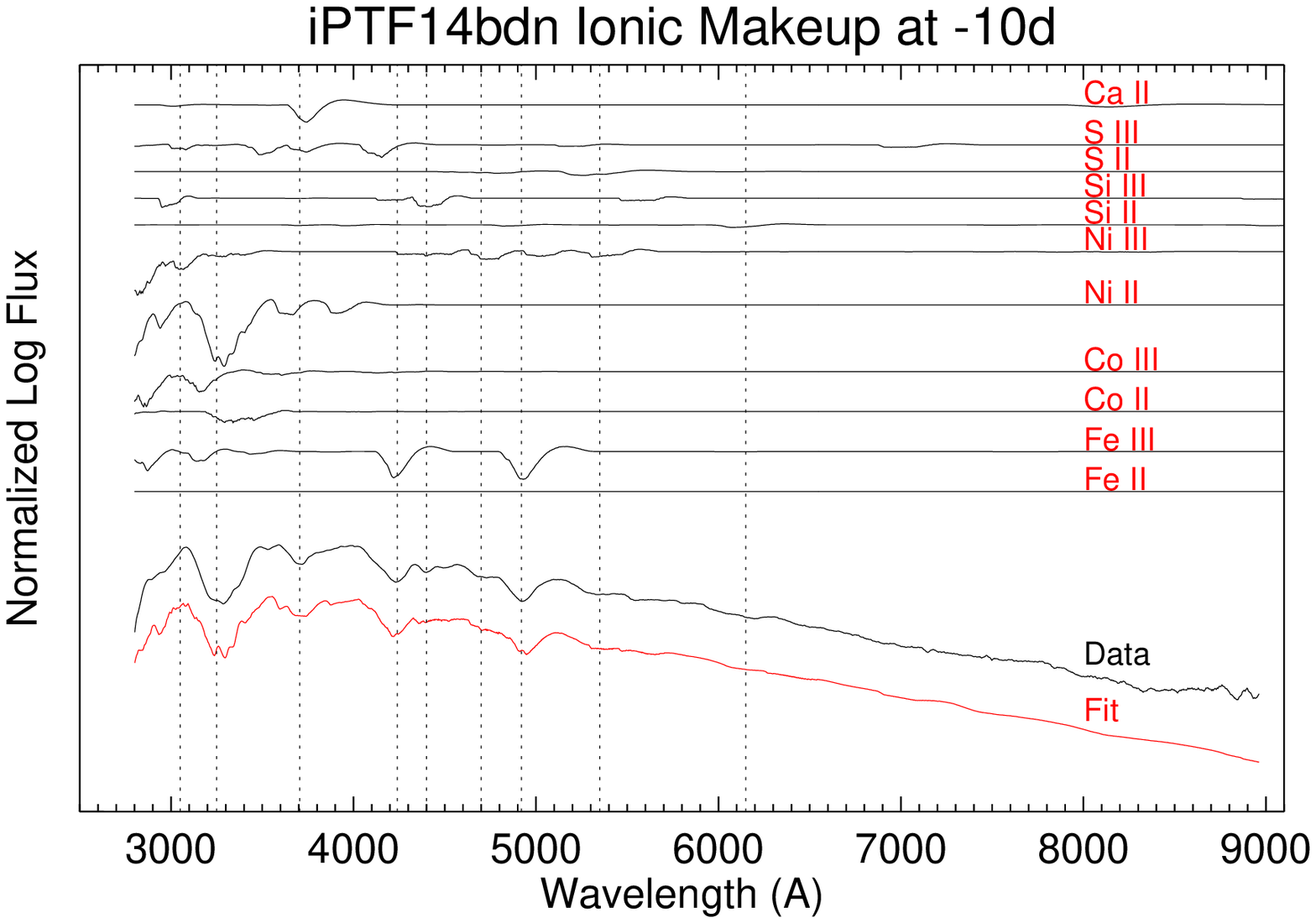}
\label{synapps-11}}
\subfigure{%
\includegraphics[height=2.8in,width=3.4in]{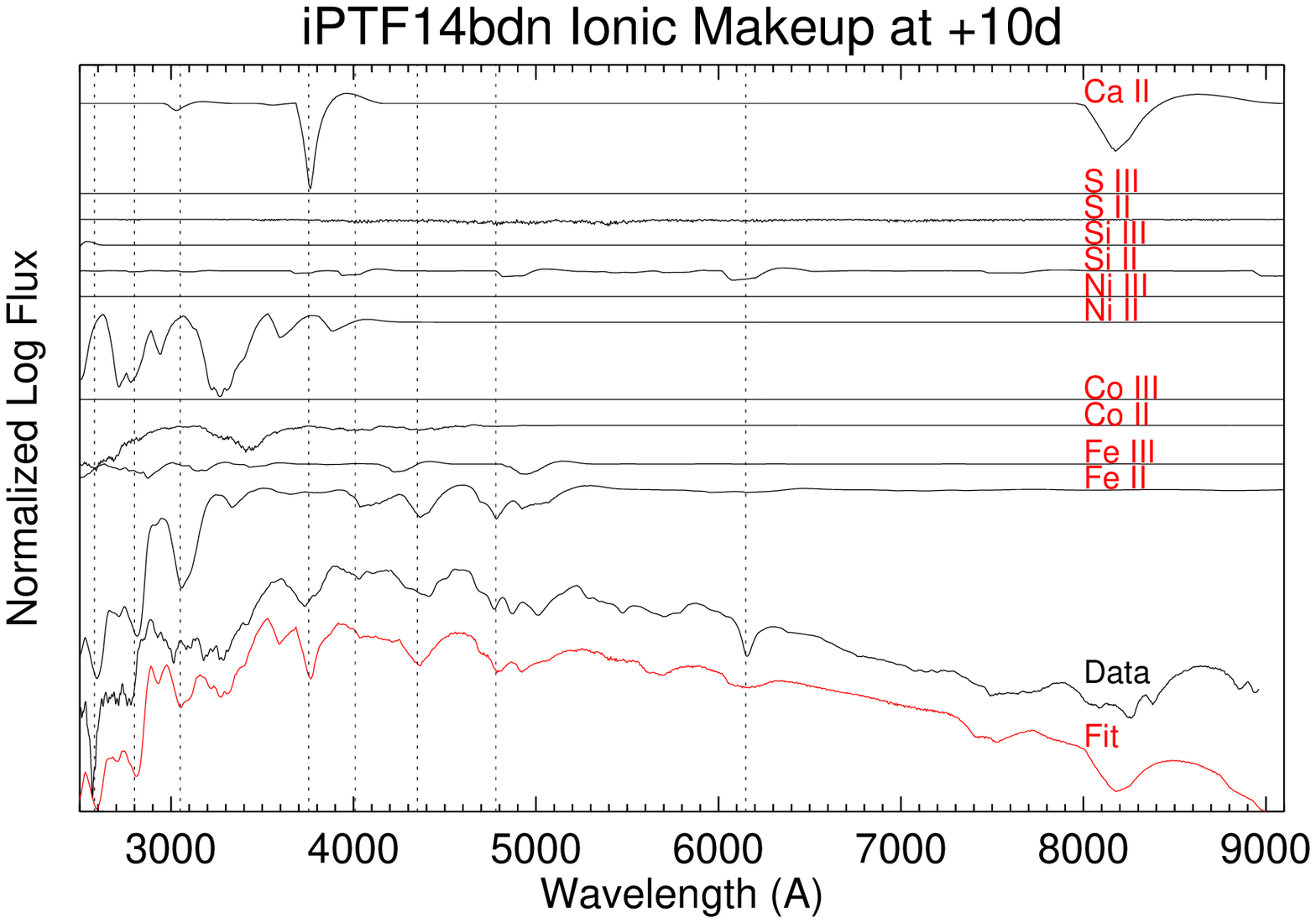}
\label{synapps+10}}
\caption{SYNAPPS fits of combined UV and optical spectra of iPTF14bdn at $-10$ and $+10$~days~ relative to B-band maximum.  The observed spectra are labeled as 'Data' and plotted in black.  SYNAPPS fits is labeled as 'Fit' and are plotted in red.  Individual ion absorption features of the SYNAPPS fits were calculated using SYN++ and are shown above the data.}
\label{figuresynapps}
\end{figure*}

To investigate the atomic structure of the SN ejecta we performed SYNAPPS model fits to the spectra near $-10 \,$ and $+10 \,$ days relative to {\it B}-band maximum light.  Timely computation of the model fits required the use of a high performace computer cluster.\footnote{The authors acknowledge the Texas A\&M University Brazos HPC cluster that contributed to the research reported here.  Brazos Computational Resource, Academy for Advanced Telecommunications and Learning Technologies, Texas A\&M University (brazos.tamu.edu).}  For this purpose the UVOT and optical spectra at the epochs nearest these phases were smoothed, normalized, and stitched together.  The model parameters assumed that all ions were attached to the photosphere and equal weight was placed on all wavelengths of the spectra.  Individual ionic contributions were generated using SYN++ and the best fit parameters computed by SYNAPPS.  The resulting model fits with the ion line strengths are presented in Figure \ref{figuresynapps}.  The model fits included all ions shown and additional contributions from Mg II, Na I and O I, which are not shown.

In the model fit of the $-10~$ day spectrum we find evidence that singly and doubly ionized iron group elements dominate the line forming regions.  The two strongest features in the optical near $4250$~\AA~ and $4910$~\AA~ are attributed to Fe III absorption.  Ni III is also identified in the optical by shallow features near $4700$~\AA~ and $5350$~\AA.  Both of these ions also display significant absorption blueward of $3500$~\AA.  Co III is shown to have little to no absorption redward of $4000$~\AA~ while contributing significantly blueward of $3300$~\AA.  Fe II and Co II are shown to produce less absorption than their doubly ionized counterparts while Ni II dominates blueward of $4000$~\AA.  Si III and S III are shown to be more prominent than their singly ionized counterparts.  Very weak Si II absorption is observed near $6100$~\AA.  A Ca II H \& K absorption feature was required by the model to fit the observed absorption feature near $3700$~\AA.  This Ca II feature is weaker than that of a normal SN Ia at this early epoch and when present with strong doubly ionized iron group elements is considered to be the hallmark of a 1999aa-like SN Ia as opposed to a 1991T-like SN Ia \citep{Garavini2004}.

At $+10$~days~ our model shows the iron group elements to be predominantly in the first ionization state.  While Fe III still shows weak absorption features, Fe II now dominates in both the optical and UV.  We find that absorption blueward of $3000$~\AA~ is predominantly the result of Fe II with Ni II contributing to a lesser degree. Our model now finds no evidence of Co III and Ni III and instead find significant absorption from Co II and Ni II.  The transition from doubly ionized iron group elements to singly ionized states is shown to now suppress the early-time feature near $3000$~\AA.  In particular, the strong Fe II absorption feature that has developed at this wavelength suggests that it is the ion primarily responsible for the evolution of this feature.  Si II is now prominent and comparable in strength to normal SNe Ia with no Si III features evident.  Ca II is now prominent near $3700$~\AA~ and $8200$~\AA~ and comparable in strength to SN 2011fe.  

\begin{figure}
\hspace*{-.2in}
\epsscale{1.2}
\plotone{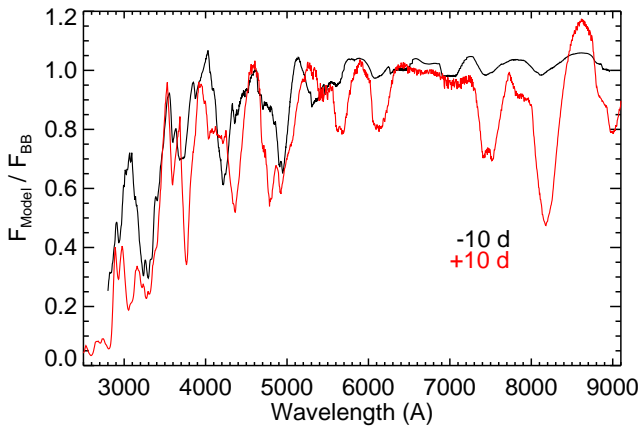}
\caption{The best-fit SYNAPPS model spectra of the data at $-10$~and $+10$~days divided by the corresponding best-fit underlying blackbody emission.  The $-10$~day data shows significant transmission of the blackbody flux between $2800-3200$~\AA, suggesting a low opacity at this epoch within this wavelength range.  We attribute this opacity window to iron group ions at higher than normal temperatures above the photosphere based on the ion analysis in Figure \ref{figuresynapps}.  This transmission decreases in the $+10$~day model.}
\label{plotopacity}
\end{figure}

In Figure \ref{plotopacity} we show the $-10$~ and $+10$~day best-fit model spectra divided by the best-fit blackbody emission profiles as calculated by SYNAPPS and SYN++.  The best-fit model photospheric temperatures were $9,200$~ and $8,800$~ K for the $-10$~ and $+10$~day spectra respectively.  This provides insight into the total effective opacity of the line forming region in the model as a function of wavelength and is useful for comparison to other models \citep{Hoeflich1993}.  In the $-10$~day model near $3000$\,\AA~ we see evidence for significant transmission of the underlying blackbody flux.  The proportion of transmitted flux is greatly reduced in the $+10$~day model.

\subsection{Comparison to 1999aa and 1991T-like SNe Ia} \label{comp91T}
\begin{figure}
\hspace*{-.2in}
\epsscale{1.2}
\plotone{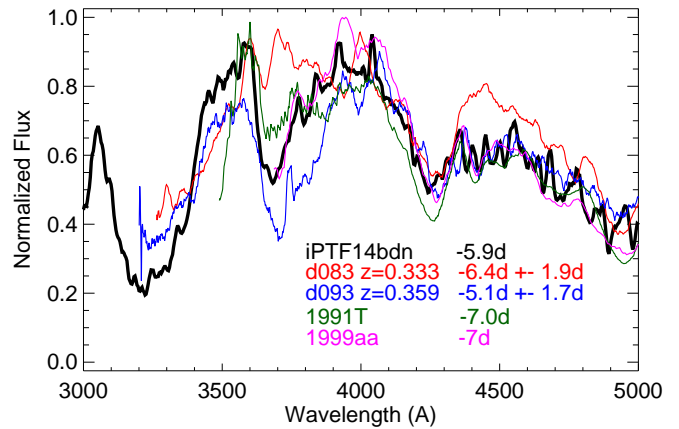}
\caption{Comparison of the rest-frame UV spectra of iPTF14bdn, SN 1991T, the ESSENCE sample of 1991T-like SNe and SN 1999aa around 6 days prior to B-band maximum light.  We observe a heterogeniety in ion features within the wavelength range of $3500-4000$~\AA~, likely due to the variable contribution of Ca II H \& K.  The presence or lack of a Ca II H \& K feature in this range is used as the classification criteria for 1999aa-like or 1991T-like SN.  We observe that iPTF14bdn is most similar to SN 1999aa.  The distribution of absorption strengths in this region suggests that these two subclasses of SNe Ia are not distinct events, but rather form a continuum.}
\label{spectra2}
\end{figure}

In Figure \ref{spectra2} we compare the rest-frame of iPTF14bdn at 6 days prior to B-band maximum light to comparable epoch spectra of 1991T \citep{Mazzali1995}, two high-z 91T-like SNe Ia from the ESSENCE Project \citep{Miknaitis2007,Foley2009} and SN 1999aa \citep{Garavini2004}.  We observe iPTF14bdn to closely resemble SN 1999aa at this epoch.  Within the regions of spectral overlap we find an overall good agreement among the spectra with the exception of the range $3500-4000$~\AA, where we observe heterogeniety.  \citet{Sasdelli2014} performed abundance tomography modeling of the spectra of SN 1991T within this wavelength range and determined that the ions which made the strongest contributions to SN 1991T at this epoch were Co III, S III or Si III with minimal contribution from Ca II H \& K.  \citet{Garavini2004} defines the presence of a Ca H \& K absorption feature in this region to be the primary distinctive characteristic between 1999aa-like and 1991T-like SNe Ia.  Among all of these SNe the strengths of this absorption feature is much weaker than that of a normal SNe Ia.

\begin{figure}
\hspace*{-.6in}
\epsscale{1.35}
\plotone{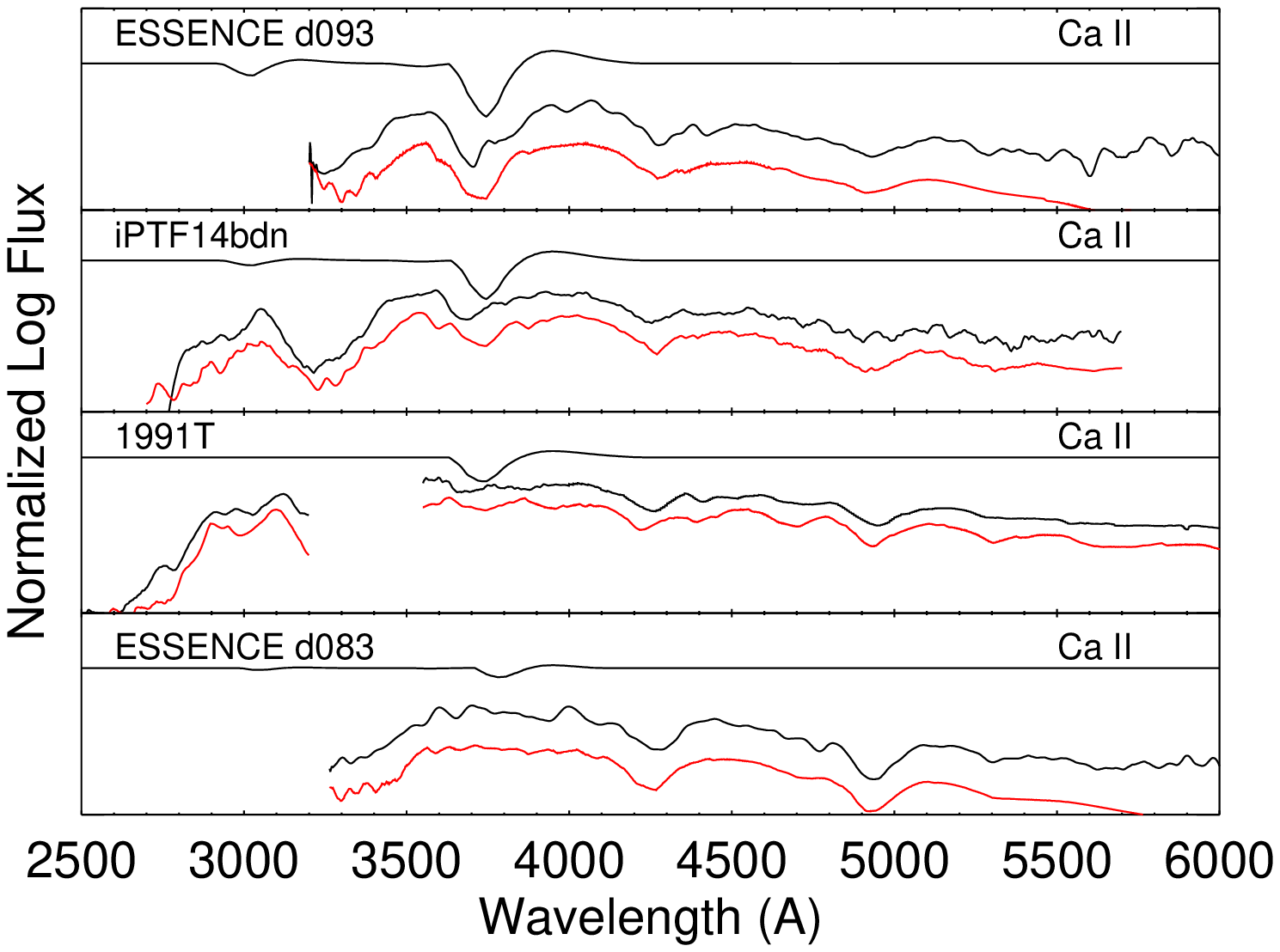}
\caption{SYNAPPS model fits to iPTF14bdn, SN 1991T, and ESSENCE SNe d083 and d093 are presented in red.  The SN 1991T spectrum is composed of {\it IUE} and ground based data.  The SYN++ model absorption profiles of the Ca II ion are plotted above each spectrum.  Ca II is found to be present in all spectra, even when not detectable by visual analysis.  The strength of the Ca II absorption is shown to be proportional to the depth of the $3500-4000$~\AA~ absorption feature.}
\label{calciumII}
\end{figure}

Among the SNe plotted we observe a continuum of absorption strengths within this wavelength range.  We computed SYNAPPS fits of the iPTF14bdn, SN 1991T and the ESSENCE SNe spectra using the same ions as our previous analysis and found that the varying strength of the absorption feature in this range corresponds to a varying contribution of Ca II H \& K.  The SYNAPPS fits and the relative strengths of the Ca II absorptions are shown in Figure \ref{calciumII}. This can be indicative that 1991T-like and 1999aa-like events are part of a continuous distribution of physical parameters.

It is noteworthy that of the five SNe we analyzed d093 of the ESSENCE project displays the strongest absorption feature in this wavelength range. The classifications of \citet{Foley2009} grouped 1991T-like and 1999aa-like SNe Ia into a single category due to their spectroscopic similarities.  The absorption feature between $3500-4000$~\AA~ and its attribution to Ca II presented here suggest that this SN is most appropriately classified as a 1999aa-like event.  


\section{Discussion} \label{discuss}

\subsection{Origin of Early-time UV Flux} \label{discussabsmags}
The analysis of SNe Ia absolute magnitudes in the UVOT bandpasses of \citet{Brown2010} included two 1991T-like SNe Ia: 2007S and 2007cq, and no 1999aa-like SNe.  In this work it was shown in the {\it uvm2} ($\lambda_{eff\_SNe\_Ia}=2360$~\AA) filter the absolute peak magnitude of 2007cq was $\sim2$~mag~ brighter than normal SNe Ia.   
The data of 2007S \citep{Milne2010} shows {\it uvw1} peaking $2.4$~days~ earlier than the B-band maximum.    

UVOT photometry has previously been presented for the 1991T-like SNe 2007S, 2007cq and 2011dn \citep{Brown2010,Milne2010,Milne2013,Brown2014a}.  UV colors are observed for 2007cq at $\sim -7~$days to $~+21~$days and 2011dn for $\sim -5~$days to $\sim +12~$days.  \citet{Milne2010} found SN 2007S to have a high reddening value of $E(B-V)~=~0.53$ and did not present data in the {\it uvm2} band, and thus we opt to not use this SN as a comparison object in this case.  In the ($uvw1~-~v$) color the three SNe trace the evolution of normal SNe Ia at epochs later than $\sim 5~$days.  In ($uvm2~-~uvw1$) iPTF14bdn and SN 2011dn show normal evolution later than $\sim 5~$days while SN 2007cq is $\sim 1~$mag bluer.  For both colors it is only in the case of iPTF14bdn that the earliest epoch photometric observations exist and show bluer than normal colors and color evolution towards redder values contrasting normal evolution.

The photometric evidence for increased UV flux at early phases is in agreement with our observation of the $2800-3200$~\AA~ feature in the earliest UV spectra of iPTF14bdn.  It is likely that this feature is the source of the greater than normal {\it uvw1} luminosity of the 1991T-like SN measured by \citet{Brown2010} and in the photometry presented here for a 1999aa-like SN.  This feature is also likely to contribute to the observed high luminosity in the {\it uvm2} band, which is moderately sensitive to this wavelength regime.  Our SYNAPPS analysis of this epoch's spectra suggests that the origin of this greater than normal luminosity is largely due to the presence of doubly ionized iron group elements which provide a low opacity window at these early epochs.  For 1999aa-like SNe we conclude that the UVOT photometric filters appear able to trace the effect of temperature evolution of the iron group elements, most notably the progression of Fe III cooling to Fe II.  Due to their spectroscopic similarities, we also expect this to be true for 1991T-like SNe Ia as well.  

\subsection{Progenitor System} \label{progenitor}
In Section \ref{synappssection} we presented evidence of Fe III, Co II, Co III, Ni II and Ni III in the $-10~$day UV-optical spectrum of iPTF14bdn.  By $+10$~days we see an evolution towards Fe II, Fe III, Co II and Ni II.  If we assume a model of expanding ejecta shells above a radiating photosphere these results suggest a continuous distribution of iron group elements in the ejecta.  In this scenario, the higher than normal temperatures at early times can be explained by the radioactive decay of greater than normal Ni ion abundances in the outer layers.  The possible physical origin of this behavior can be an explosion mechanism that produces significant mixing of newly synthesized iron group elements into the outer layers.  

\subsection{Origin of the I-band Secondary Maximum}
Secondary maxima in infrared light curves are often observed in SNe Ia.  Models of the physical origins of these features have  suggested that the effect is dependent upon the mass and distribution of $^{56}$Ni in the ejecta, and upon the recombination of doubly ionized iron group ions \citep{Hoeflich1995,Kasen2006}.  The analysis of \citet{Jack2015} identified the secondary maximum in the {\it I}-band to be the result of Fe II emission from high excitation transitions following recombination of doubly ionized elements to singly ionized states.  Their conclusion was that the {\it I}-band captures a transient Fe II emission feature at $\sim 7500$~\AA~ which temporarily increases the in-band flux.  Evidence supporting this conclusion was presented using optical spectra of the normal SNe 2014J and 2011fe just before and during the secondary maximum and \uppercase{phoenix} models at similar epochs.  They didn't make any investigation of the presence of this feature in peculiar SNe Ia.

\begin{figure}
\epsscale{1.2}
\plotone{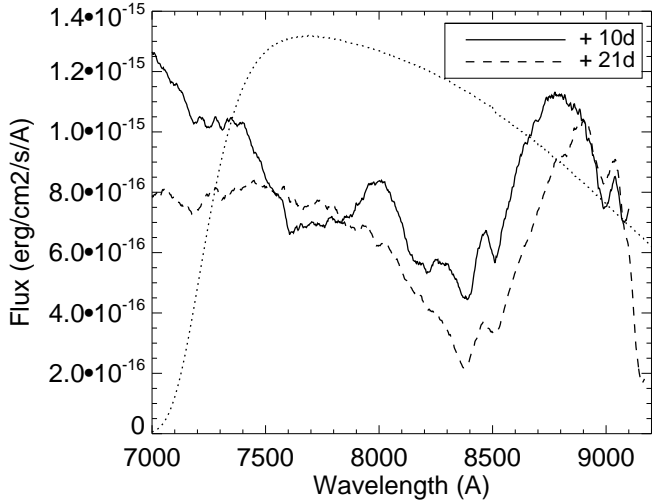}
\caption{Spectra of iPTF14bdn at phases prior to and during the {\it I}-band secondary maximum.  The dotted curve represents the normalized YFOSC {\it I}-band filter transmission function.  We see evidence of an increase in flux at the epoch of {\it I}-band secondary maximum near $\sim 7500$~\AA.  In the normal SNe Ia 2014J and 2011fe this feature was attributed to recombination of Fe III to Fe II and was identified as the source of the {\it I}-band secondary maximum by \citet{Jack2015}.}
\label{Ibump}
\end{figure}

The photometry of iPTF14bdn in Figure \ref{lightcurves} displays a secondary maximum in the {\it I}-band occurring around $20$~days after B-band maximum light.  Our identification in Sec. \ref{synappssection} of Fe III recombining to Fe II between $-10$ and $+10$~days, the persistent presence of Fe III at $+10$~days, and the well documented similarities of normal, 1991T-like and 1999aa-like SNe at these epochs suggest that the secondary maximum may be caused by the same feature as \citet{Jack2015}.  Our spectroscopic sample enables us to investigate the occurrence of this phenomenon in our data set by making a comparison of the optical spectral features just prior to and during the {\it I}-band secondary maximum.  In Figure \ref{Ibump} we compare the optical spectra of June 24 and July 5, at phases $+10$ and $+21$~days respectively, in the region of the Lijiang $2.4$~m's YFOSC {\it I}-band sensitivity.  The combined transmission efficiency of the YFOSC {\it I}-band filter and CCD quantum efficiency is shown as a dotted line to highlight the spectral region to which our photometric observations are most sensitive.  We show that the photometric system used is most sensitive to $\sim 7700$~\AA~ and is thus well suited for detecting the $7500$~\AA~ Fe II emission noted by \citet{Jack2015}.  

In iPTF14bdn we observe a similar increase in flux around $\sim 7500$~\AA~ as \cite{Jack2015} at the epoch of {\it I}-band secondary maximum.  This confirms that the same physical processes of the normal SNe 2014J and 2011fe are likely at work in this 1999aa-like SN.  We conclude that the source of the {\it I}-band secondary maximum in iPTF14bdn is likely attributable to the same Fe II emission as in normal SNe.

We note that here the comparison spectra come from epochs of roughly equal {\it I}-band magnitude, the earlier epoch being about $5$~days prior to the {\it I}-band trough minimum.  This is in slight contrast to those of \citet{Jack2015}, whose pre-secondary maximum spectra are timed closer to the trough minimum and lower fluxes.  Thus, we expect our flux differences to be less stark, as is the case.  

Prompted by our detection of Fe III recombining to Fe II between $-10$~ and $+10$~days, we investigated the possibility of detecting the same Fe II emission feature in our optical spectra of $-13$, $-11$, and $-5$~days by comparing to SN 2011fe \citep{Mazzali2014,Pereira2013}.  In all cases we found no evidence for Fe II emission around $\sim 7500$~\AA.

\section{Conclusion} \label{conclusion}
We have presented UV and optical photometry and spectra of the 1999aa-like SN iPTF14bdn.  Our photometric observations show slowly declining, broad lightcurves and pre-maximum color evolution from blue to red UV colors at epochs earlier than $-5$~days.  This behavior is in contrast to normal SNe Ia, indicating a hotter than normal environment at early times.  

We presented the first UV spectra series of a 1999aa-like SN as well as optical spectra spanning the epochs $-15$ to $+40$~days relative to {\it B}-band maximum light.  Comparison of the UV spectra to two normal SNe Ia revealed greater than normal flux at $-9$~days in the regions of the Ca II H \& K feature and blueward of $3200$~\AA.   iPTF14bdn is shown to progress with time toward normal spectral features and becomes indistinguishable from normal SNe around one week following {\it B}-band maximum, as is typical in the optical for 1991T-like and 1999aa-like SNe.  

Using SYNAPPS models matched to combined UV and optical spectra of iPTF14bdn at $-10$ and $+10$~days relative to {\it B}-band maximum light we showed that the spectra progress over time from being dominated by doubly ionized ions to singly ionized species, notably Si, S, Fe, Co and Ni.  Ion analysis of the model fits show the observed early-time blue UV photometric colors can be attributed to an opacity 'window' blueward of $3200$~\AA~ resulting from iron group elements occupying doubly ionized states.  With time the iron group elements cool and recombine to singly ionized states and the opacities begin to resemble those of normal SNe Ia.  These UV spectroscopic features can be explained by higher than normal temperatures at early times resulting from a high Ni mass mixed into the outer ejecta layers.  

Spectral comparison to a sample of 1991T-like and 1999aa-like SNe Ia around $6$~days prior to {\it B}-band maximum showed that iPTF14bdn most closely resembles SN 1999aa.  
In the $3500~-~4000$~\AA~ range we see a continuous distribution of absorption features in the spectra attributed to Ca II absorption in SYNAPPS model fits.  We conclude that the 1991T-like and 1999aa-like subclassifications of SNe Ia may not be distinct subclasses, but rather comprise a single class with varying amounts of Ca II absorption.

Finally, we determined that the source of the {\it I}-band secondary maximum can likely be attributed to Fe II emission around $7500$~\AA.  We identified an increase in flux around $7500$~\AA~ at the epoch of {\it I}-band secondary maximum using the method of \citet{Jack2015}, who identified this phenomenon in the normal SNe Ia 2014J and 2011fe. 

\acknowledgements
The authors would like to thank Paul Kuin for the UVOTPY software used in extracting the UVOT spectra, Rollin Thomas for the SYN++ and SYNAPPS spectra synthesis packages, and Chris Burns for the SNooPy light curve analysis program.  We also wish to thank the George P. and Cynthia Woods Mitchell Institute for Fundamental Physics \& Astronomy at Texas A\&M and the Carnegie Supernova Project.  Michael Smitka is supported partially by the NASA Swift Guest Investigator grant NNX15AR51G.
P. J. Brown and the Swift Optical/Ultraviolet Supernova Archive are supported by NASA's Astrophysics Data Analysis Program through grant NNX13AF35G.  We acknowledge the support of the staff of the Li-Jiang 2.4-m telescope (LJT), and Tsinghua-NAOC 80-cm telescope (TNT). Funding for the LJT has been provided by Chinese academy of science (CAS) and the People's Government of Yunnan Province.   The TNT is owned by Tsinghua University and operated by the National Astronomical Observatory of the Chinese Academy of Sciences (NAOC). 
Ju-Jia Zhang is supported by the National Natural Science Foundation of China (NSFC, grant 11403096). The work of Xiao-Feng Wang is supported by the Major State Basic Research Development Program (2013CB834903), the NSFC (grants 11073013, 11178003, 11325313), Tsinghua University Initiative Scientific Research Program, and the Strategic Priority Research Program ``The Emergence of Cosmological Structures" of the Chinese Academy of Sciences (grant No. XDB09000000).
This research has made use of the NASA/IPAC Extragalactic Database (NED) which is operated by the Jet Propulsion Laboratory, California Institute of Technology, under contract with the National Aeronautics and Space Administration.

\bibliographystyle{apj}
\bibliography{bibtex_14bdn}

\appendix \label{appendix}
In order to facilitate precision comparison of our photometry to other photometric systems we provide the photometric transmission functions of the YFOSC instrument in Table \ref{yfosc}.  Transmission values for each filter were calculated by multiplying the CCD response curve into the filter transmission curves.  The transmission values are given in units of percent of photons detected.  The photometric transmission functions of the {\it Swift} UVOT filters are maintained by NASA's HEASARC and are available at http://heasarc.gsfc.nasa.gov/docs/heasarc/caldb/data/swift/uvota/index.html.

\begin{table}[h]
	\begin{center}
	\caption{YFOSC Instrument Filter Responses}	
	\label{yfosc}
	\begin{tabular}{ccc}
	\toprule\toprule
	Filter & Wavelength & Throughput\\
	         & (\AA) & (\%)\\
	\midrule
	U  & 3130 &  0.026 \\
	U  & 3132 &  0.028 \\
	U  & 3134 &  0.034 \\
	U  & 3136 &  0.042 \\
	U  & 3138 &  0.049 \\
	U  & 3140 &  0.055 \\
	U  & 3142 &  0.064 \\
	U  & 3144 &  0.073 \\
	U  & 3146 &  0.084 \\
	U  & 3148 &  0.094 \\
	\bottomrule
	\end{tabular}
	\end{center}
\end{table}
\begin{center}
This table is available in its entirety in machine-readable form in the online journal.
\end{center}

\end{document}